# Long Tails, Automation and Labor

B.N. Kausik[1]

July 16 2023

## Abstract

A central question in economics is whether automation will displace human labor and diminish standards of living. Whilst prior works typically frame this question as a competition between human labor and machines, we frame it as a competition between human consumers and human suppliers. Specifically, we observe that human needs favor long tail distributions, i.e., a long list of niche items that are substantial in aggregate demand. In turn, the long tails are reflected in the goods and services that fulfill those needs. With this background, we propose a theoretical model of economic activity on a long tail distribution, where innovation in demand for new niche outputs competes with innovation in supply automation for mature outputs. Our model yields analytic expressions and asymptotes for the shares of automation and labor in terms of just four parameters: the rates of innovation in supply and demand, the exponent of the long tail distribution and an initial value. We validate the model via non-linear stochastic regression on historical US economic data with surprising accuracy.

JEL D63, E22, E23, E24, J24, O33, O41

[1] Conflict disclosure: unaffiliated independent.
Author's bio: https://www.linkedin.com/in/bnkausik/  Contact: bnkausik@gmail.com
Thanks to D. Autor, F. Baskett, J. Jawahar, R. Krishnan, K. Suresh and A. Salomons.

# Introduction

We consider the question whether automation will displace human labor and diminish standards of living. Prior work, such as Frey and Osborne (2013), Korinek and Stiglitz (2020), Brynjolfsson (2022) and Acemoglu (2022) are pessimistic, while others such as Arntz, Gregory and Zierahn (2017), Bessen (2019), Autor et al (2020), and Basu (2022) offer reasons for optimism.

One perspective is that human labor and automation compete for their respective share of economic activity, e.g. Acemoglu and Restrepo (2018), Acemoglu and Restrepo (2019), Autor et al (2022), and that automation inexorably wins, driving down labor's share of economic output, Karabarbounis and Neiman (2014), Oberfield and Raval (2014). Some of these works view the competition between labor and automation through the lens of the impact of cost of capital on automation. However, cost of capital affects both consumption and investment in automation, and although cost of capital and Labor Share of Gross Domestic Income are positively cross correlated, cost of capital is negatively cross correlated with Median Real Wages as in Fig. 1 (data source: St. Louis Federal Reserve).

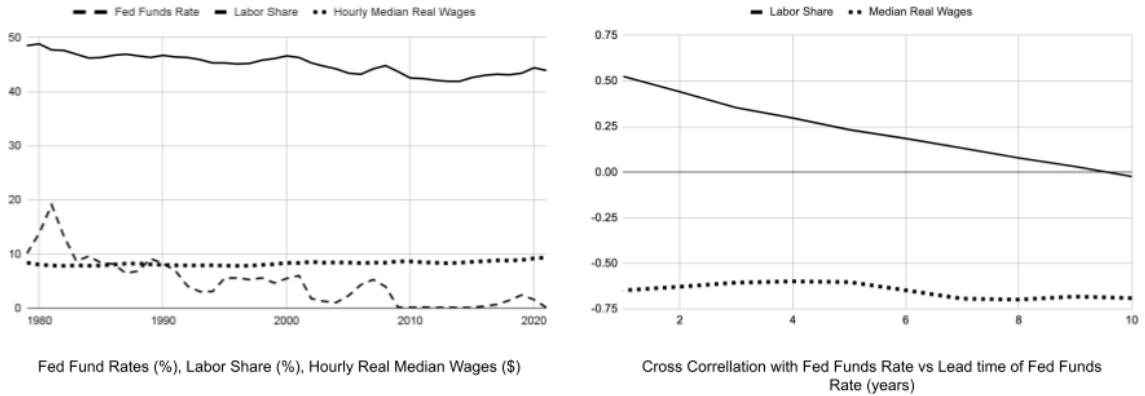

Fig. 1: Fed Funds Rate, Labor Share and Median Real Wages

Our approach is as follows:
1. We observe that human needs favor long tail distributions, demanding the same from goods and services that fulfill those needs.
2. We propose a theoretical model of economic activity on a long tail distribution, where innovation in demand for output variants on the long tail competes with innovation in supply automation at the head of the distribution. In other words, on the demand side, human consumers innovate in seeking new niche variants of goods and services. While on the supply side, human producers innovate on automation to drive down the cost of mature goods and services. Our model yields analytic expressions and asymptotes for the shares of automation and labor in terms of just four parameters: the rates of innovation in supply and demand, the exponent of the long tail distribution and an initial value.
3. We validate our model on historical US economic data via non-linear stochastic regression on two measures of labor share and estimate the asymptotes.

# The Long Tail of Human Needs

The conventional wisdom of the so-called 80/20 rule, also known as the Pareto principle, is that 80% of the results accrue from 20% of the causes. The 20% is the "head" of the distribution, and the remaining 80% of the distribution is the "long tail" with a weak return on effort and not worth pursuing. However, when it comes to human needs and tastes, the Pareto principle seems at best transient. For example, during the early years of the mass market automobile, the Ford Model T enjoyed a market share of almost 50%. However, as the market expanded and matured, competitors emerged to cater to the fragmented diversity of human tastes and needs. As of 2022, the top 20% of automobile brands represent ~60% of vehicles sold in the US market, see Fig. 1.

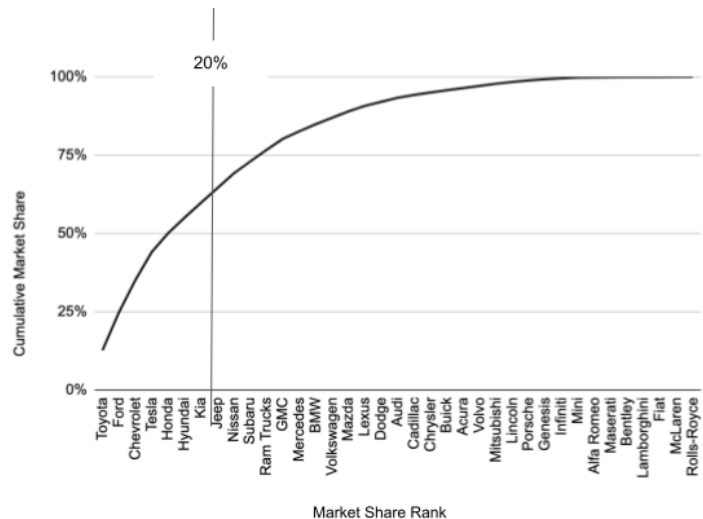

Fig. 2: US automobile market (2022)

With the advent of the internet, it became economically feasible to extend the long tail in many areas of human interest. Anderson (2004) recognized the importance of the long tail, and thereafter Anderson (2006) and Brynjolfsson et al (2007) expanded on the ideas in the context of retail. Goel et al (2010) present an analysis of the long-tail in retail, music, movies and web search, and observe that small increases in covering the long tail can result in disproportionately large increases in user satisfaction. Kausik (2023) builds on Goel et al (2010) to reason that automation must perform on the long tail to displace human labor in white collar service occupations.

# The Long Tail of Economic Activity

Based on our discussion in the previous section, we propose a theoretical macroeconomic model where economic activity follows a long tail distribution. Referring to Fig. 3, the horizontal axis is all available tasks in the economy ranked in decreasing order of economic output, where the rank fraction of a task is the ratio of the rank of the task over the total number of tasks. The vertical axis is the cumulative economic output share $P(r)$ of tasks with rank fraction between 0 and $r$. At any point in time, there exists rank fraction $a$ such that tasks in $[0, a]$ are performed by automation while tasks in $(a, 1]$ are performed by labor. We refer to $a$ as the automation fraction.

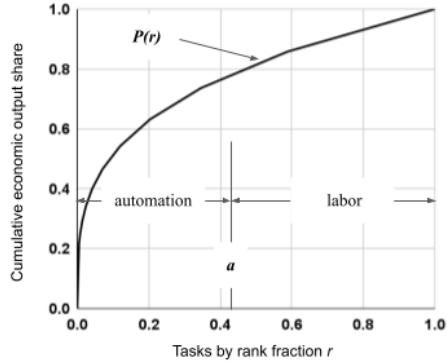

Fig. 3. Long tail model of economy

**Assumption 1:** The cumulative economic output share is given by $P(r) = r^n$ where $n$ is a constant model parameter. Automation's share of the output is $a^n$ and labor's share is $(1 - a^n)$.

**Assumption 2:** Tasks leftmost in the labor component are newly automated at a nominal rate $\sigma \geq 0$, in that the automation fraction $a$ nominally increases to $a + \sigma\tau$ after infinitesimal time interval $\tau$. See Fig. 4. We refer to $\sigma$ as the *supply innovation rate*.

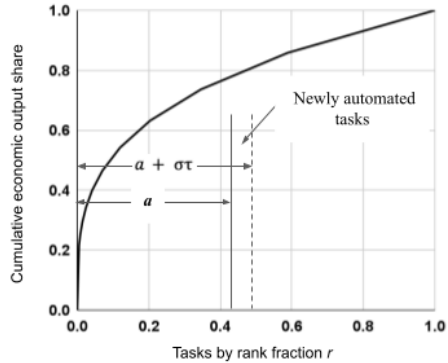

Fig. 4. Supply innovation: expanded automation of tasks

**Assumption 3:** As existing tasks become obsolete and new niche tasks are created, the rank fraction of tasks improves at a nominal rate $\delta > 0$, in that each point $r$ on the horizontal axis nominally shifts to $r(1 - \delta\tau)$ after infinitesimal time interval $\tau$. See Fig. 5. We assume $\sigma/\delta \leq 1$ is a constant, and refer to $\delta$ as the *demand innovation rate*.

Our motivation with Assumptions 2 and 3 is to allow for the rates of innovation in supply and demand to vary with the cost of capital, whilst preserving their ratio. If the cost of capital is low/high, innovation in automation increases/decreases. Likewise, if the cost of capital is low/high, demand for output variants increases/decreases as well. As we will observe later, the ratio reflects the demographics of the economy.

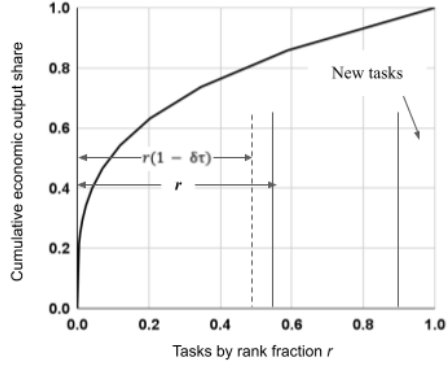

Fig. 5. Demand innovation: obsolescence and new niche tasks

**Theorem 1:** The automation fraction $a$ is determined by

$$a = \frac{\sigma}{\delta} + \left(a_0 - \frac{\sigma}{\delta}\right) e^{-\int_{t_0}^{t} \delta \, dt}$$

where $a_0$ is the automation fraction at initial time $t_0$.

**Proof:** Combining Assumptions 2 and 3, we get

$$da/dt = \sigma - a\delta.$$

Rearranging, we get

$$da/dt = -\delta\left(a - \frac{\sigma}{\delta}\right)$$

Thence

$$\frac{1}{\left(a - \frac{\sigma}{\delta}\right)} da = -\delta \, dt$$

Per Assumption 3, $\sigma/\delta$ is a constant. Hence integrating both sides and setting $a = a_0$ at $t = t_0$, we get

$$a = \frac{\sigma}{\delta} + \left(a_0 - \frac{\sigma}{\delta}\right) e^{-\int_{t_0}^{t} \delta \, dt}$$

**Corollary 1:** The automation fraction $a$ is asymptotically $\frac{\sigma}{\delta}$.

**Proof:** From Theorem 1,

$$\lim_{t \to \infty} a = \lim_{t \to \infty} \left[ \frac{\sigma}{\delta} + \left(a_0 - \frac{\sigma}{\delta}\right) e^{-\int_{t_0}^{t} \delta dt} \right] = \frac{\sigma}{\delta} \text{ since } \delta > 0 \text{ and } \sigma/\delta \text{ is a constant.}$$

**Corollary 2:** Labor's share of the output is

$$1 - \left[ \frac{\sigma}{\delta} + \left(a_0 - \frac{\sigma}{\delta}\right) e^{-\int_{t_0}^{t} \delta dt} \right]^n$$

**Proof:** Per Assumption 1, labor's share of the output is $(1 - a^n)$. Substituting for $a$ per Theorem 1, the result follows.

**Corollary 3:** Labor's share of the output is asymptotically $1 - \left(\frac{\sigma}{\delta}\right)^n$.

**Proof:** Per Assumption 1, labor's share of the output is $(1 - a^n)$. Substituting for $a$ per Corollary 1, the result follows.

# Validation

We now validate our model against historical economic data.

**Data Set 1:** We consider the St. Louis Federal Reserve's measure entitled "Shares of gross domestic income: Compensation of employees, paid: Wage and salary accruals: Disbursements: to persons (W270RE1A156NBEA)" reported annually for the period 1948 through 2021. As a simplification, we assume $\delta$ is a constant and run non-linear regression with TensorFlow on the 74 data points to extract the four parameters in the labor share formula of Corollary 2 as below:

$$1 - \left[ \frac{\sigma}{\delta} - K e^{-\delta(t-t_0)} \right]^n$$

Each run consists of 100 iterations of stochastic gradient descent and Mean Square Error loss, starting with the four parameters randomly initialized in the interval [0,1]. We average the four parameters across 100 runs to obtain the following values:

$$n = 8.68 \times 10^{-2}; \; K = 1.56 \times 10^{-1}; \; \delta = 1.27 \times 10^{-2}; \; \sigma = 7.39 \times 10^{-3}$$

Fig. 6 plots the results. The thick gray line is the actual historical data across the regression period 1948-2021. The thin line is a linear regression of the actual historical data. The solid black line is the model's fit across the same period. The dotted line is the model's prediction beyond. The Root Mean Squared error of the model in the regression period is 2.4%. And per Corollary 3, the asymptotic limit of the labor share is ~37%

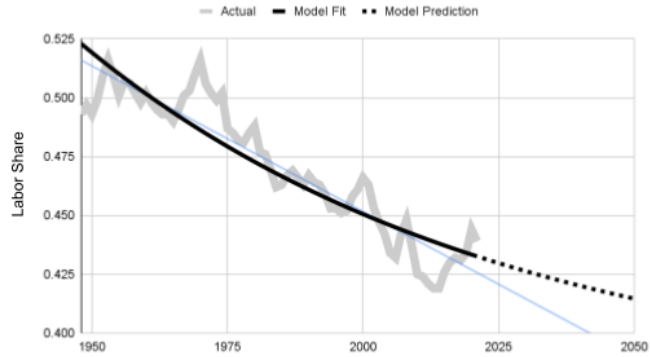

Fig. 6. Labor share from Gross Domestic Income

**Data Set 2:** The data set based on gross domestic income with annual samples has too much noise and too few data points to be processed effectively as two spans, one for regression and one for testing the model. Towards that goal, we synthesize an estimate as follows:

$$Labor\ Share\ =\ \frac{Real\ Median\ Wages\ \times Labor\ Participation\ Rate\times (1-Unemployment)}{Real\ Per\ Capita\ GDP}$$

The measures on the right hand side of the formula are available quarterly from the St. Louis Federal Reserve between Jan 1979 and Mar 2023 for a total of 177 data points:
- Median usual weekly real earnings: (LES1252881600Q)
- Labor Force Participation Rate (CIVPART)
- Real gross domestic product per capita (A939RX0Q048SBEA)
- Unemployment Rate (UNRATE)

We split the 177 data points into two equal spans, Jan 1979 to Mar 2001, and Jun 2001 to Mar 2023, and extract the model parameters via stochastic regression on the first span as below.

$$n = 8.57 \times 10^{-1};\ K = 5.80 \times 10^{-1};\ \delta = 2.17 \times 10^{-2};\ \sigma = 1.7 \times 10^{-2}$$

We then use the model to predict labor share in the second span. Fig. 7 plots the results. The thick gray line is the actual historical data between Jan 1979 and Mar 2023. The thin line is a linear regression of the actual historical data. The solid black line is the model fit across the regression period Jan 1979 to June 2001. The dotted line is the model's prediction for the test period June 2001 to Mar 2023, and beyond. The Root Mean Squared error of the model is 1.5% and 3.4% in the regression and test periods respectively. And per Corollary 3, the asymptotic limit of the labor share is ~19%.

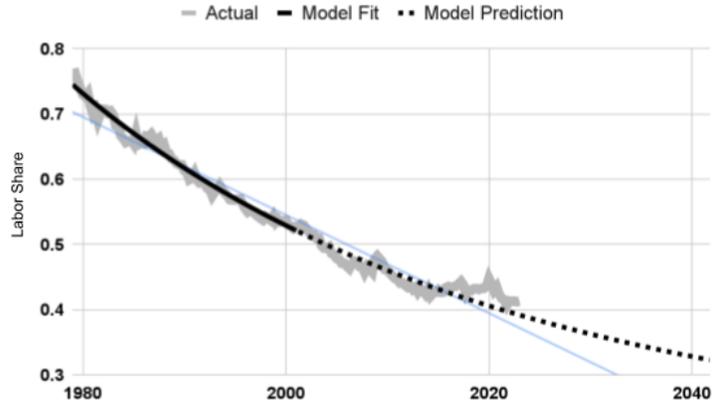

Fig. 7. Labor share from Real Median Wages

We now ask whether σ/δ is indeed a constant per Assumption 3. One possibility is to view the number of people employed in science & technology jobs as a proxy for the Supply Innovation Rate σ, and the number of college educated people in the population as a proxy for the Demand Innovation Rate δ as below:

$$\sigma/\delta \sim \frac{Number\ Employed\ in\ Science\ \&\ Technology}{Number\ of\ College\ Educated\ in\ Population}$$

Fig. 8 shows the above ratio to be largely flat between 1960-2017, (data source: NSF & Census).

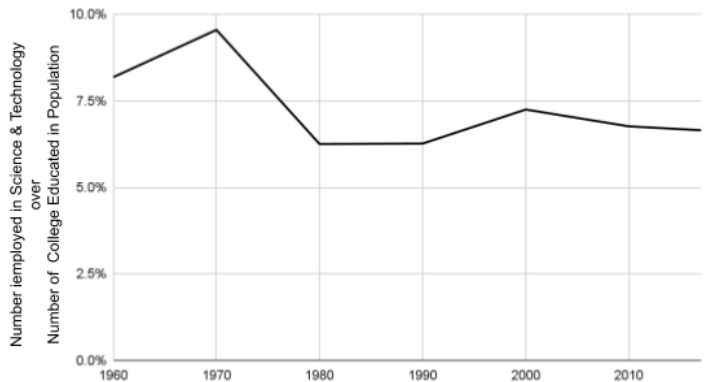

Fig. 8. Estimated ratio of Supply Innovation Rate and Demand Innovation Rate

# Summary

A central question in economics is whether automation will displace human labor and diminish standards of living. Whilst prior works typically frame this question as a competition between human labor and machines, we frame it as a competition between human consumers and human suppliers. Specifically, we observe that human needs favor long tail distributions, i.e., a long list of niche items that are substantial in aggregate demand. In turn, the long tails are reflected in the goods and services that fulfill those needs. With this background, we propose a theoretical model of economic activity on a long tail distribution, where innovation in demand for new niche outputs competes with innovation in supply automation for mature outputs. Our model yields analytic expressions and asymptotes for the shares of automation and labor in terms of just four parameters: the rates of innovation in supply and demand, the exponent of the long tail distribution and an initial value. We validate the model via non-linear stochastic regression on historical US economic data with surprising accuracy.